\input harvmac
\def\frak#1#2{{\textstyle{{#1}\over{#2}}}}
\def\frakk#1#2{{{#1}\over{#2}}}

\def\pa{\partial}
\def\semi{;\hfil\break}

\def\DRED{\ifmmode{{\rm DRED}} \else{{DRED}} \fi}
\def\DREDp{\ifmmode{{\rm DRED}'} \else{${\rm DRED}'$} \fi}
\def\NSVZ{\ifmmode{{\rm NSVZ}} \else{{NSVZ}} \fi}
  
\def\npb{{Nucl.\ Phys.\ }{\bf B}}   
\def\prd{{Phys.\ Rev.\ }{\bf D}}

\def\plb{{Phys.\ Lett.\ }{\bf B}}

\def\Ycal{{\cal Y}}   
\def\Ocal{{\cal O}}
\def\Qcal{{\cal Q}}
\def\Rcal{{\cal R}}
\def\btY{\overline{\tilde Y}{}}
\def\btmu{\overline{\tilde \mu}{}}
   
\def\betabar{\overline\beta{}}
\def\gammabar{\overline\gamma{}}
\def\thbar{{\overline\theta}{}}
\def\sbar{\overline\sigma{}}
\def\Fbar{\overline F{}}
\def\Wbar{\overline W{}}

\def\Phbar{\overline\Phi{}}

\def\mbar{{\overline m}{}}
     
\def\bhat{\hat\beta{}}
\def\bbhat{\hat{\overline\beta}{}}

\def\bbar{{\overline b}{}}
\def\cbar{{\overline c}{}}
\def\hbar{{\overline h}{}}
{\nopagenumbers
\line{\hfil LTH 496} 
\line{\hfil hep-ph/0103255}
\vskip .5in
\centerline{\titlefont Gauge Singlet Renormalisation}
\centerline{\titlefont in Softly-Broken Supersymmetric Theories}
\vskip 1in
\centerline{\bf I.~Jack, D.R.T.~Jones and R.~Wild}
\medskip
\centerline{\it Dept. of Mathematical Sciences,
University of Liverpool, Liverpool L69 3BX, UK}
\vskip .3in
We consider the renormalisation of a  softly-broken supersymmetric
theory with singlet fields and a superpotential with  a linear term.   We
show that there exist exact $\beta$-functions for both the linear term
in the superpotential and  the associated linear term in the Lagrangian.
We also construct  exact renormalisation group invariant trajectories
for these terms, corresponding to the    conformal anomaly solution for
the soft masses and couplings.

\Date{March 2001}
In supersymmetric gauge theories which contain gauge singlet fields (such as 
the MSSM with the addition of right-handed neutrinos) there
is the possibility of a linear term in the superpotential $W$,
so that we have (for a renormalisable theory) 
\eqn\suppot{W(\phi) = a^i\phi_i 
+ \frak{1}{2}{\mu}^{ij}\phi_i\phi_j + \frak{1}{6}Y^{ijk}
\phi_i\phi_j\phi_k.}
In the component 
formalism the $a^i$ term leads to a term in the
scalar potential which is linear in the auxiliary field 
$F$. 
In some of our previous work on the renormalisation group functions of 
softly-broken 
supersymmetric theories\ref\jja{I.~Jack and D.R.T.~Jones,
\plb415 (1997) 383}\nref\jjpa{I.~Jack, D.R.T.~Jones and A.~Pickering,  
\plb 426 (1998) 73}%
\nref\jjpb{I.~Jack, D.R.T.~Jones and A.~Pickering, 
\plb432 (1998) 114}%
--\ref\jjpc{I.~Jack, D.R.T.~Jones and A.~Pickering, 
\plb435 (1998) 61}, we have excluded  singlets and hence 
such terms; so  our purpose here is to extend 
the formalism to incorporate them. 

The renormalisation issues raised by a linear $F$-term are similar to those 
associated with a Fayet-Iliopoulos (FI) linear $D$-term, 
which is possible when the gauge group contains an abelian factor.  
In previous papers\ref\xius{I.~Jack and D.R.T.~Jones, \plb473 (2000) 
102}\nref\jjp{I.~Jack, D.R.T.~Jones and S.~Parsons, \prd62 (2000) 125022}%
--\ref\jjxib{I.~Jack, D.R.T.~Jones, \prd63 (2001) 075010}
we computed the $\beta$-function for the coefficient $\xi$ of this term,
and showed that upon eliminating $D$ using its equation of motion, 
$\beta_{\xi}$ is associated with additional terms in the $\beta$-function for 
the soft masses. We also showed the existence of a solution to the
renormalisation group (RG) equations for $\xi$, related to the exact anomaly 
mediated supersymmetry breaking (AMSB) solutions for the soft breaking 
parameters\ref\con{L. Randall and R. Sundrum,  \npb 557 (1999) 79\semi
G.F. Giudice, M.A. Luty, H. Murayama and  R. Rattazzi,
JHEP 9812 (1998) 27
}\ref\conus{I. Jack and D.R.T.~Jones, \plb 465 (1999) 148},
but which in this case could only be constructed order by order in
perturbation theory.  In the present paper we shall perform the
analogous analysis  for the linear $F$-term. In this case the
$\beta$-function for $a^i$ is associated, after elimination 
(or, as we shall see, redefinition) of $F$, with
additional terms in the  $\beta$-functions for the quadratic and linear
soft scalar couplings. Therefore the treatment of a linear $F$-term
involves both generalising our previous analysis  of the quadratic soft
term and a discussion of the linear soft term. 

The analysis will be simpler than the $D$-term 
case, since (by superspace power counting in the spurion formalism) 
$a^i$ can only receive 
logarithmically divergent corrections, 
whereas the individual diagrams contributing to $\beta_{\xi}$ are quadratically 
divergent, and  so although this quadratic divergence cancels 
when the graphs are 
summed,  the evaluation of an individual contribution to $\beta_{\xi}$ 
in the spurion formalism 
is non-trivial.  By contrast, in the linear $F$ case 
the full power of the spurion formalism may be 
brought to bear, 
leading to exact results for the relevant 
$\beta$ functions\foot{This analysis was to an extent anticipated
in Ref.~\ref\yam
\ref\yam{Y.~Yamada, \prd50 (1994) 3537}.} 
and corresponding exact AMSB solutions. For pedagogical reasons, 
however, we will begin in the component formalism.

The scalar potential of our component Lagrangian is 
\eqn\vpot{V = V_{\rm susy} + V_{\rm soft},}
where
\eqn\vsusy{
V_{\rm susy} = - F^i F_i - F^i \frakk{\pa W^*}{\pa \phi^i}
-F_i \frakk{\pa W}{\pa \phi_i} - \frak{1}{2}(D^a)^2 - gD^a \phi^* R^a \phi,}
and
\eqn\vsoft{
V_{\rm soft} = \left(c^i\phi_i
+\frak12b^{ij}\phi_i\phi_j+\frak16h^{ijk}\phi_i\phi_j\phi_k
-\kappa^i{}_jF^j\phi_i+\hbox{c.c.}\right) +(m^2)^i{}_j\phi_i\phi^j,}
where as usual $\phi^i = (\phi_i)^*$, and the 
supermultiplet $(\phi,F,\psi)$ transforms 
according to the representation $R^a$.  
We have included the standard soft-breaking terms together with the
additional terms involving $c$ and $\kappa$ required for multiplicative 
renormalisability. Notice that although we have a $F^*\phi$ term in Eq.~\vsoft,
we do not add a $F\phi$ one because the latter would lead (in general) to 
quadratically divergent tadpoles; for the same reason there is 
no $\phi^*\phi^2$ term. 

It is a simple matter to show that if we define 
\eqn\fdefn{
\Fbar_i = F_i + \kappa^j{}_i\phi_j + a_i}
then we obtain 
\eqn\newV{\eqalign{V &= - \Fbar^i \Fbar_i 
- \Fbar^i \frakk{\pa \Wbar^*}{\pa \phi^i}
-\Fbar_i \frakk{\pa \Wbar}{\pa \phi_i} - \frak{1}{2}(D^a)^2 
- gD^a \phi^* R^a \phi\cr
& +\left(\cbar^i\phi_i 
+ \frak{1}{2}\bbar^{ij}\phi_i\phi_j 
+ \frak{1}{6}\hbar^{ijk}\phi_i\phi_j\phi_k +\hbox{c.c.}\right)
+(\mbar^2)^i{}_j\phi_i\phi^j,\cr}} 
where
\eqn\newW{
\Wbar(\phi) = \frak{1}{2}{\mu}^{ij}\phi_i\phi_j + \frak{1}{6}Y^{ijk}
\phi_i\phi_j\phi_k,}
and
\eqna\redef$$\eqalignno{
\left(\mbar^2\right)^i{}_j &= \left(m^2\right)^i{}_j
+\left(\kappa\kappa^{\dagger}\right)^i{}_j, &\redef a\cr
\hbar^{ijk} &= h^{ijk}+Y^{l(jk}\kappa^{i)}{}_l,&\redef b\cr
\bbar^{ij} &= b^{ij}+Y^{ijl}a_l+\mu^{l(i}\kappa^{j)}{}_l,&\redef c\cr
\cbar^i &= c^i+ \mu^{il}a_l+\kappa^i{}_l a^l,&\redef d\cr}$$
with
$$Y^{l(jk}\kappa^{i)}{}_l = Y^{ljk}\kappa^i{}_l+Y^{ilk}\kappa^j{}_l
+Y^{ijl}\kappa^k{}_l
$$
and
$$\mu^{l(i}\kappa^{j)}{}_l = \mu^{li}\kappa^j{}_l+\mu^{lj}\kappa^i{}_l.$$
The relations Eqs.~\redef{a-d} are renormalisation group invariant,
because  Eq.~\vpot\ has all the interactions necessary for 
multiplicative renormalisability.  Although $\kappa$ and $a$ are 
necessary for this RG invariance, it is clear that they are not
independent  couplings, since they do not appear in the reduced
potential,  Eq.~\newV. We have only examined the  scalar sector above,
but note that we can simply replace $W$ by $\Wbar$ in the  fermion
sector, since this depends on the second derivatives of $W$ with
respect to $\phi$. The equivalence of a
theory with a superpotential like $W$ to  one with a superpotential like
$\Wbar$  might seem puzzling, since in a theory with no soft terms, 
linear terms are  a {\it sine qua non\/} for $F$-type spontaneous supersymmetry
breaking\ref\lian{L.~O'Raifeartaigh, \npb 96 (1975) 331}.   The
resolution lies, of course, in the $Y^{ijl}a_l$ and $\mu^{il}a_l$ terms
in Eq.~\redef{c,d} respectively, which are generated by the redefinition
 of $F$. 

We now essentially present the above analysis again, but using the spurion 
formalism, which, allied with the non-renormalisation theorem, will enable 
us to derive a series of exact relations among the $\beta$-functions.

In the spurion context, the Lagrangian corresponding to Eqs.~\vsusy,
\vsoft\ is given by
\eqn\lagf{
L = L_{\rm susy}+L_{\rm soft}+L_{\rm GF}+L_{\rm FP},}
where
\eqn\lsusy{
L_{\rm susy}=\int d^4\theta \Phbar^j\left(e^{2gV}\right){}^i{}_j\Phi_i
+\left[\int d^2\theta 
\left(W(\Phi)+\frak14 W^{\alpha}W_{\alpha}\right)+\hbox{c.c.}\right],}
and
\eqn\lsoft{\eqalign{
L_{\rm soft}=&-\left[\int d^2\theta \theta^2
\left(c^i\Phi_i +\frak{1}{2}b^{ij}\Phi_i\Phi_j
+ \frak{1}{6}h^{ijk}\Phi_i\Phi_j\Phi_k
+\frak{1}{2}MW^{\alpha}W_{\alpha}\right)
+\hbox{c.c.}\right]\cr
&+\int d^4\theta \left[-(m^2)^k{}_j\theta^2\thbar^2\Phbar^j
\left(e^{2gV}\right){}^i{}_k\Phi_i+\Phbar^j\left(\theta^2\kappa^k{}_j+\thbar^2
\kappa^{\dagger k}{}_j
\right)\left(e^{2gV}\right){}^i{}_k\Phi_i\right],\cr}}
where $V$ is the vector superfield, $W^{\alpha}$
the corresponding field strength and $M$ is the gaugino mass. 
$L_{\rm GF}$ and $L_{\rm FP}$ are the 
gauge-fixing and ghost Lagrangians whose exact form will not be important to
us. Now by making the redefinition 
\eqn\redefa{
\Phi_i = \Phi'_i-\theta^2(\kappa^j{}_i\Phi'_j+a_i),}
(which corresponds precisely to Eq.~\fdefn) we find that the Lagrangian adopts 
the form 
\eqn\lagfa{
L' = L'_{\rm susy}+L'_{\rm soft}+L_{\rm GF}+L_{\rm FP},}
where
\eqn\lsusya{
L'_{\rm susy}
= \int d^4\theta \Phbar^{\prime j}\left(e^{2gV}\right){}^i{}_j\Phi'_i
+\left[\int d^2\theta \left(\Wbar(\Phi') + 
\frak{1}{4} W^{\alpha}W_{\alpha}\right)+\hbox{c.c.}\right],}
and
\eqn\lsofta{\eqalign{
L'_{\rm soft}=&-\left[\int d^2\theta \theta^2
\left(\cbar^i\Phi'_i +\frak{1}{2}\bbar^{ij}\Phi'_i\Phi'_j 
+ \frak{1}{6}\hbar^{ijk}\Phi'_i\Phi'_j\Phi'_k
+ \frak{1}{2}
MW^{\alpha}W_{\alpha}\right)
+\hbox{c.c.}\right]\cr
&-\int d^4\theta (\mbar^2)^k{}_j\theta^2\thbar^2\Phbar^{\prime j}
\left(e^{2gV}\right){}^i{}_k\Phi'_i,\cr}}
where $\hbar$, $\bbar$, $\cbar$ and $\mbar^2$ are exactly as defined in 
Eq.~\redef{}, and $\Wbar$ as defined in Eq.~\newW. 
Once again, note that $a$ and $\kappa$ no longer appear explicitly. 
We shall refer to $L$ in Eq.~\lagf\ as the unreduced
Lagrangian, and $L'$ in Eq.~\lagfa\ as the reduced Lagrangian; it is 
the latter that one would use in practical applications.   

We may now obtain a set of consistency conditions by requiring $L$ and $L'$ in 
Eqs.~\lagf, \lagfa\ to be equivalent as functions of the renormalised 
couplings (i.e. equal for all renormalisation scales $\mu$). 
We use $\betabar$, $\beta$ to represent a $\beta$-function calculated in the 
 reduced,  unreduced formalisms respectively. Moreover, for each
$\beta$ function we separate out the part $\bhat$ corresponding to 
1PI graphs. For example, we write 
\eqn\hatunhat{
\beta^a_i(a,b,\cdots) = \gamma^m{}_ia_m+\bhat^a_i
,}
where $\bhat^a = \bhat^a (Y,Y^*,g,b,\mu,M,h^*)$ is determined
by 1PI tadpole graphs.
Writing 
\eqn\gamrel{
\mu{d\over{d\mu}}\Phi'_i = -\gamma'^j{}_i\Phi'_j
=-\gamma^j{}_i\Phi'_j-2\theta^2\gammabar^j_{1i}\Phi'_j
+\theta^2\sbar_i,}
it follows from RG invariance of Eq.~\redefa\ that 
\eqn\bdel{
\gammabar_1^i{}_j = -\frak12\bhat_{\kappa}^i{}_j
-\kappa^i{}_k\gamma^k{}_j,}
and 
\eqn\cons{
\sbar_i = \bhat^a_i+2\gamma^m{}_ia_m.}
Eqs.~\bdel\ and \cons\ give $\gammabar_1^i{}_j$ and $\sbar_i$ in terms of the
unreduced parameters; we shall shortly give prescriptions for calculating them
directly in terms of reduced parameters.

From the non-renormalisation theorem we have
\eqn\betay{
\beta_Y^{ijk} = Y^{l(jk}\gamma^{i)}{}_l,}
with a similar expression for $\beta_{\mu}$.  However, as foreshadowed in 
Eq.~\hatunhat, in the presence of soft terms the non-renormalisation
theorem does not protect $\beta_a$ from 1PI contributions, 
and $\bhat_a$ is non-zero.  
We do, however, have 
\eqn\betahu{
\beta_h^{ijk} = h^{l(jk}\gamma^{i)}{}_l,}
with a similar result for $\beta_b$; but again $\bhat_c$ is non-zero. 
To derive the soft $\beta$-functions in the reduced formalism, we 
impose 
\eqn\rgl{
\mu{d\over{d\mu}}\left(L_{\rm susy}+L_{\rm soft}\right) = \mu{d\over{d\mu}}
\left(L'_{\rm susy}+L'_{\rm soft}\right).}
Then using the results for the unreduced $\beta$-functions such as 
Eqs.~\betay, \betahu, inserting Eq.~\gamrel\ and using Eqs.~\bdel\ and \cons,
we obtain    
\eqna\barbetas$$\eqalignno{
\betabar_{\hbar}^{ijk} &= \hbar{}^{l(jk}\gamma^{i)}{}_l -
2Y^{l(jk}\gammabar_1{}^{i)}{}_l, &\barbetas a\cr
\betabar_{\bbar}^{ij} &= 
\bbar{}^{l(i}\gamma^{j)}{}_l-2\mu{}^{l(i}\gammabar_1{}^{j)}{}_l
+Y^{ijl}\sbar_l, &\barbetas b\cr}$$
together with consistency conditions relating reduced and unreduced
quantities (analogous to Eqs.~\bdel\ and \cons)
\eqn\ccons{   
\bbhat_{\cbar}^i=\bhat^i_{c}-2a^l\left(\gammabar_1\right){}^i{}_l+\bhat_a^l
\kappa^i{}_l+\mu^{il}\sbar_l,}
and
\eqn\mcons{
\left(\bbhat_{\mbar^2}\right){}^i{}_j = \left(\bhat_{m^2}\right){}^i{}_j
-2\left(\kappa\gammabar_1^{\dagger}\right){}^i{}_j
-2\left(\gammabar_1\kappa^{\dagger}\right){}^i{}_j
-2\left(\kappa\gamma\kappa^{\dagger}\right){}^i{}_j.}
Eqs.~\barbetas{}--\mcons\ may also be obtained (perhaps more simply) 
by operating with $\mu{d\over{d\mu}}$ on the RG-invariant
relations Eqs.~\redef{}. 
Eqs.~\barbetas{}\ were given in Refs.~\jja\ref\akk{L.V.~Avdeev, 
D.I.~Kazakov and I.N.~Kondrashuk,
\npb510 (1998) 114}, but excluding singlet fields. 
Results in the presence of
singlet fields were given up to the two-loop level in 
\yam\ref\jjold{I. Jack and D.R.T.~Jones, \plb 333 (1994) 372}\ref\mandv{S.P. 
Martin and  M.T. Vaughn, \prd 50 (1994) 2282}.

We now consider the explicit forms of $\gammabar_1$, $\betabar_{\mbar^2}$,
$\sbar$ and 
$\bbhat_{\cbar}$. 
From Eq.~\gamrel, we see that $\gammabar_1$ and $\sbar$ are obtained from 
$\theta^2$-dependent contributions to the two-point function and the 
one-point function respectively, calculated from $L'$. We also see from 
Eq.~\lsofta, on rewriting 
\eqn\rewrit{
\int d^2\theta \theta^2 c^i\Phi_i = \int d^4\theta \theta^2 \thbar^2c^i\Phi_i,}
that $\betabar_{\mbar^2}$ and $\betabar_{\cbar}$ are obtained from the
$\theta^2\thbar^2$-dependent contributions to the two-point function and the
one-point function respectively, again calculated from $L'$.
We are led to the following prescription: consider superspace diagrams 
contributing to the two-point function in the supersymmetric theory, with 
superpotential $\Wbar$. 
These diagrams are only logarithmically 
divergent, and so for the purposes of the $D$-algebra the 
$\theta^2, \thbar^2$ associated with the spurion 
may be taken as constants. In each
such diagram  we can simply
replace $Y^{ijk}$ by $Y^{ijk}-\hbar^{ijk}\theta^2$, $\mu^{ij}$ by 
$\mu^{ij}-\bbar^{ij}\theta^2$, and gauge couplings $g^2$ by
by $g^2(1+M\theta^2+M^*\thbar^2+MM^*\theta^2\thbar^2)$\yam. 
We also replace
each factor $\delta^k{}_l$ in an internal chiral propagator by
$\delta^k{}_l+(\mbar^2)^k{}_l\theta^2\thbar^2$. This procedure may be 
implemented using differential operators; for instance,
we obtain for $\gammabar_1$ and $\betabar_{\mbar^2}$
\eqna\Odef$$\eqalignno{
(\gammabar_1)^i{}_j  &= {\cal O}\gamma^i{}_j, &\Odef a\cr
\left(\betabar_{\mbar^2}\right)^i{}_j &= \Delta\gamma^i{}_j, &\Odef b\cr}$$
where
\eqna\Otdef$$\eqalignno{
{\cal O}  &= Mg^2\frakk{\partial}{\partial g^2}-\hbar^{lmn}
\frakk{\partial}{\partial Y^{lmn}}-\bbar^{lm}\frakk{\pa}{\pa \mu^{lm}},
&\Otdef a\cr
\Delta &= 2\Ocal\Ocal^* +2MM^* g^2{\partial
\over{\partial g^2}} 
+\left[\btY^{lmn}{\partial\over{\partial Y^{lmn}}}
+\btmu^{lm}{\partial\over{\partial \mu^{lm}}}+ \hbox{c.c.}\right]
+X{\partial\over{\partial g}},&\Otdef b}$$
with
\eqn\tydef{
\btY^{ijk} = (\mbar^2)^{(i}{}_lY^{jk)l} \quad\hbox{and}\quad
\btmu^{ij} = (\mbar^2)^{(i}{}_l\mu^{j)l}.}
Eq.~\Odef{a}\ was given in Refs.~\jja,\akk; note however the inclusion 
of the derivatives with respect to $\mu$, which give zero acting 
on $\gamma$ but will be important presently. The full understanding
of Eq.~\Odef{b}, in particular the necessity for, and form of, the term
involving $X$ in Eq.~\Otdef{b},
was developed in Refs.~\jja, \jjpb\ and also Refs.~\akk,
\ref\vk{D.I.~Kazakov and V.N.~Velizhanin, \plb485 (2000) 393} (see also 
Ref.~\ref\arkh{G.F. Giudice and  R. Rattazzi, \npb 511 (1998) 25\semi
N.~Arkani-Hamed  et al, \prd 58 (1998) 115005}). In 
particular,  it was shown in Ref.~\jjpb\ that
a form for $X$ derived for a particular RG trajectory in Ref.~\ref\kkz{
T.~Kobayashi, J.~Kubo, and G.~Zoupanos,
\plb 427 (1998) 291} was in fact valid in general. (Note that the $X$ term,
hitherto written separately, has now been included in the definition of 
$\Delta$.)

For $\sbar$ and $\bbhat_{\cbar}$ we
should consider superspace tadpole diagrams. By chirality, such diagrams can be
obtained by taking a graph contributing to the two-point function with an 
external leg attached to a $Y$ or $Y^*$, and replacing this
$Y$ or $Y^*$ by a $\mu$ or $\mu^*$ 
respectively. After making the substitutions described above,
$\sbar$ and $\bbhat_{\cbar}$ are derived from the $\theta^2$ terms 
and $\theta^2\thbar^2$ terms respectively in these tadpole diagrams.
This process may be accomplished using the operators 
${\cal O}$ and $\Delta$ defined above. We obtain
\eqn\bsup{
\sbar_i = -2{\cal O}\left( Z_i\right),}
where 
\eqn\Zdef{
Z_i = Y_{imn}K^{mn}{}_{pq}\mu^{pq},}
with $K^{mn}{}_{pq}$ defined by the condition
\eqn\Xdef{
Y_{imn}K^{mn}{}_{pq}Y^{pqj}a_j = \gamma^j{}_ia_j.}
We also have
\eqn\betac{
\bbhat^i_{\cbar} = \Delta Z^i+\mu^{il}\sbar_l-(\mbar^2)^i{}_kZ^k.}

These results
hold in the reduced case; however, the calculations of $\bhat^a$
and $\bhat_c$ in the unreduced
case are closely related, corresponding to $\theta^2$ terms
and $\theta^2\thbar^2$ terms respectively in tadpole diagrams derived from
$L$ rather than $L'$. The difference is that there are now 
contributions from insertions of $\kappa$ on internal propagators,
which correspond to substituting for 
$\hbar$, $\bbar$, $\cbar$, $\mbar^2$ in terms of $h$, $b$, $c$, $m^2$ in 
accordance with Eq.~\redef{}.
However, making these substitutions in Eqs.~\bsup, \betac\ overcounts by 
including contributions (from $a$, and from insertions of $\kappa$ on
external legs) which would correspond to one-particle reducible
diagrams. This leads precisely to the
consistency conditions Eqs.~\cons, \ccons. Similar reasoning applies
when considering the relation between $\gammabar_1$ and $\bhat_{\kappa}$.
In this case the
substitution of $\hbar,\ldots$ in terms of $h,\ldots$ does not yield all 
the $\kappa$ contributions to   
$\bhat^i_{\kappa j}$, which also contains a contribution 
$\kappa^i{}_k[Y^{kmn}K_{mn}{}^{pq}Y_{pqj}-\gamma^k{}_j]$, with $K$ as in 
Eq.~\Xdef.  
However, the substitution overcounts by 
including $-\kappa^i{}_kY^{kmn}K_{mn}{}^{pq}Y_{pqj}$ which would correspond to
one particle reducible diagrams and must be removed. 
Combining these contributions leads to Eq.~\bdel.  

The results Eqs.~\bsup\ and \betac\ mean that our knowledge of $\sbar$ 
and $\bbhat_{\cbar}$ is limited only by our knowledge of $\gamma$. Thus
all the  $\beta$-functions  that depend on soft-breaking parameters are
determined by the underlying  supersymmetric theory, except for the one
associated with a  FI-term. We have verified Eq.~\bsup\ by an explicit 
calculation of $F$-tadpole diagrams through three loops, using  the
Feynman gauge component formalism  and supersymmetric dimensional
regularisation.

We may now obtain exact solutions of the RG equation for $a_i$ 
and $c^i$ (in the  
 unreduced case), or 
equivalently exact solutions to the RG equations for $\bbar^{ij}$ and $\cbar^i$
(in the 
 reduced case). It is already well-known\con\conus\ that 
the following set of equations provide an exact solution (the AMSB solution) 
to the
renormalisation group equations for $M, \hbar, \bbar$ and $\mbar^2$ in the
case where there are no singlet fields and the gauge group contains no
abelian factors:
\eqna\result$$\eqalignno{M &= M_0{\beta_g\over g}, &\result a\cr
\hbar^{ijk} &=-M_0\beta_Y^{ijk},&\result b\cr
\bbar^{ij} &=-M_0\beta_{\mu}^{ij}, &\result c\cr
(\mbar^2)^i{}_j &= \frak{1}{2}|M_0|^2\mu{d\gamma^i{}_j\over{d\mu}}.
&\result d\cr}$$
In fact these solutions are realised if the only source of
supersymmetry breaking is the conformal anomaly, when $M_0$ is
the gravitino mass\con.
However, 
Eq.~\result{d}\ acquires extra terms\xius\ if the gauge group 
contains abelian factors via non-zero FI terms, 
and (as we shall show)
Eq.~\result{c}\ acquires extra terms if there are singlet fields in the theory.

In the  unreduced case, the solutions corresponding to Eq.~\result{}\ are
\eqna\resultn$$\eqalignno{
M &= M_0{\beta_g\over g}, &\resultn a \cr
h^{ijk} &= 0, &\resultn b\cr
b^{ij} &= 0, &\resultn c\cr
(m^2)^i{}_j &= |M_0|^2\left[\frak{1}{2}\mu{d\gamma^i{}_j\over{d\mu}}
-(\gamma^2)^i{}_j\right], &\resultn d\cr
\kappa^i{}_j &= -M_0\gamma^i{}_j.&\resultn e\cr}$$
RG invariance of Eqs.~\resultn{b,c} follows trivially from Eq.~\betahu\ 
and the corresponding equation for $\beta_b$;  
RG invariance of Eq.~\resultn{e} follows from Eq.~\bdel, using the 
fact that that on the AMSB 
trajectory, 
\eqn\Otraj{
\gammabar_1^i{}_j = \frak12M_0\mu{d\over{d\mu}}\gamma^i{}_j,}
a relation established in Ref.~\conus. Finally,  
the RG-invariance of Eq.~\resultn{d}\ then follows 
from that of Eq.~\result{d}\ using Eqs.~\redef{a}, \resultn{e}.

We now claim that solutions to the RG equations for $a_i$, $c^i$ 
corresponding to Eqs.~\resultn{}\ are
\eqna\atraj$$\eqalignno{
a_i &= -M_0Z_i,&\atraj a\cr
c^i &= \frak12 |M_0|^2\left[\mu{d\over{d\mu}}Z^i-(\gamma Z)^i
\right],&\atraj b\cr}$$
with $Z$ as defined in Eq.~\Zdef.
It is straightforward to show that this works; let us begin with Eq.~\atraj{a}. 
Eq.~\bsup\ now becomes simply
\eqn\bsupnew{
\sbar_i = {2\over{M_0}}\Ocal a_i,}
where, on applying Eqs.~\redef{}\ and \resultn{}\ in Eq.~\Otdef{a}, we find 
\eqn\calOb{
\Ocal = \frak{1}{2}M_0\left(\beta_g{\pa\over{\pa g}}+2\Qcal\right)
-Y^{klm}a_m{\pa\over{\pa\mu^{kl}}},}
with
\eqn\defq{
\Qcal = \sum_{klm}\beta_Y^{klm}{\pa\over{\pa Y^{klm}}}
+\sum_{kl}\beta_{\mu}^{kl}{\pa\over{\pa \mu^{kl}}}.}
On the other hand, 
\eqn\muderiv{
\mu{d\over{d\mu}} = \beta_g{\pa\over{\pa g}}+\Rcal,}
where
\eqn\calQ{
\Rcal = \Qcal+\Qcal^*.}
Now in Ref.~\conus\ it was shown that 
for a tensor $X^i{}_j$ we have
\eqn\master{   
(\Qcal X)^i{}_j-(\Qcal^* X)^i{}_j= \gamma^i{}_k X^k{}_j- X^i{}_k\gamma^k{}_j,}
and in particular that 
\eqn\mgamma{\Qcal\gamma = \Qcal^*\gamma.}
Eq.~\mgamma, in fact, is the result one needs to establish Eq.~\Otraj.
The generalisation of Eq.~\master\ to a tensor with an arbitrary number 
of indices is obvious; but for our purposes all we need is the result  
\eqn\mast{
(\Qcal Z)_i-(\Qcal^* Z)_i = -\gamma^j{}_iZ_j.}
Armed with this equation and 
\eqn\new{
Y^{klm}a_m{\pa\over{\pa\mu^{kl}}}a_i = -M_0\gamma^k{}_ia_k,}
(which follows easily from Eqs.~\atraj{a}, \Zdef, \Xdef), 
we can show (using Eq.~\cons, \bsupnew--\new) that 
\eqn\bsupa{
\bhat_i^a = \mu{d\over{d\mu}}a_i-\gamma^m{}_ia_m,}
reproducing Eq.~\hatunhat, and thereby proving the RG invariance of 
Eq.~\atraj{a}.
We now turn to Eq.~\atraj{b}. To prove RG invariance of this solution, we 
require two identities, generalising similar results proved in Ref.~\conus. The 
first (which follows by repeated application of Eq.~\mast) is 
\eqn\rsq{\eqalign{
\Rcal^2Z^i=&\Bigl(4\beta_Y^{klm}\beta^Y_{pqr}{\pa^2\over{\pa Y^{klm}\pa 
Y_{pqr}}}
+4\beta_Y^{klm}\beta^{\mu}_{pq}{\pa^2\over{\pa Y^{klm}\pa \mu_{pq}}}\cr
&+4\beta_{\mu}^{kl}\beta^Y_{pqr}{\pa^2\over{\pa \mu^{kl}\pa Y_{pqr}}}
+4\beta_{\mu}^{kl}\beta^{\mu}_{pq}{\pa^2\over{\pa \mu^{kl}\pa \mu_{pq}}}\cr
&+(\Rcal\gamma)^{(k}{}_nY^{lm)n}{\pa\over{\pa Y^{klm}}}
+(\Rcal\gamma)^n{}_{(k}Y_{lm)n}{\pa\over{\pa Y_{klm}}}\cr
&+(\Rcal\gamma)^{(k}{}_n\mu^{l)n}{\pa\over{\pa \mu^{kl}}}             
+(\Rcal\gamma)^n{}_{(k}\mu_{l)n}{\pa\over{\pa \mu_{kl}}}\Bigr)Z^i+(\gamma^2Z)^i
.\cr}}
The second identity is that if Eqs.~\result{}\ are imposed, then
\eqn\Xident{
|M_0|^2\mu{d\beta_g\over{d\mu}} = 3{\beta_g^2|M_0|^2\over{g}}+2X.}
(Note that this identity is true for a range of regularisation schemes which
includes standard dimensional reduction\jjpb.)  
Using these identities in conjunction with Eqs.~\muderiv, \Otdef{b}, \calOb,
\atraj{a} and \new, it follows that when Eqs.~\result{}\ are imposed, we have 
\eqn\deltraj{
\Delta Z^i = \frak12|M_0|^2\left[\left(\mu{d\over{d\mu}}\right)^2Z^i- 
(\gamma^2Z)^i+2\mu{d\gamma^i{}_k\over{d\mu}}Z^k\right].}
Using Eqs.~\ccons, \bsupa, \atraj{a}, \deltraj, \result{d},  
\resultn{d}, \betac, we find
\eqn\ctraj{
\beta_c^i = \frak12|M_0|^2\left[\left(\mu{d\over{d\mu}}\right)^2Z^i-
\mu{d\over{d\mu}}
(\gamma Z)^i\right],}
which shows that Eq.~\atraj{b}\ is RG-invariant.

With the aid of Eqs.~\resultn{}\ and \atraj{} it is now straightforward to 
write down the AMSB results in the reduced case, by substituting in 
Eq.~\redef{}. One can also check that the resulting expressions are indeed 
RG invariant.  For convenience we first  assemble 
the complete results for the soft $\beta$-functions 
in the reduced formalism:
\eqn\allbetas{\eqalign{
\betabar_M &= 2\Ocal\left[\frakk{\beta_g}{g}\right],\cr 
\betabar_{\hbar}^{ijk} &= \hbar{}^{l(jk}\gamma^{i)}{}_l -
2Y^{l(jk}\gammabar_1{}^{i)}{}_l, \cr
\betabar_{\bbar}^{ij} &= 
\bbar{}^{l(i}\gamma^{j)}{}_l-2\mu{}^{l(i}\gammabar_1{}^{j)}{}_l
+Y^{ijl}\sbar_l,\cr
\betabar^i_{\cbar} &= \cbar^j \gamma^i{}_j
+\Delta Z^i+\mu^{il}\sbar_l-(\mbar^2)^i{}_kZ^k,\cr
\left(\betabar_{\mbar^2}\right){}^i{}_j &= \Delta\gamma^i{}_j,\cr}}
where $\sbar$, $Z^i$ are  defined in Eqs.~\bsup, \Zdef,
and then we give the full AMSB solutions:
\eqn\amsball{\eqalign{M &= M_0{\beta_g\over g},\cr
\hbar^{ijk} &= -M_0\beta_Y^{ijk}, \cr
\bbar^{ij} &= -M_0\beta_{\mu}^{ij} - M_0Y^{ijk}Z_k,\cr
\cbar^i &= \frak12|M_0|^2
\left[\mu{d\over{d\mu}}Z^i+(\gamma Z)^i\right]-M_0\mu^{il}Z_l,\cr
(\mbar^2)^i{}_j &= \frak{1}{2}|M_0|^2\mu{d\gamma^i{}_j\over{d\mu}}.\cr
}}
For a $U_1$ theory with a FI term, 
the AMSB  solution for $\mbar^2$ in the 
$D$-eliminated case becomes\xius 
\eqn\massrg{
(\mbar^2)^i{}_j = \frak{1}{2}|M_0|^2\mu{d\gamma^i{}_j\over{d\mu}}
+g\xi^{\rm RG}(\Ycal)^i{}_j,} 
where $\xi^{\rm RG}$ is the RG solution for $\xi$, and $\Ycal$ is the 
hypercharge matrix for the $U_1$ factor, with gauge coupling $g$. 
The proof
relies on the consistency condition Eq.~(2.25) of Ref.~\xius, which plays a 
similar r\^ole to that of Eq.~\cons. 

In conclusion: we have extended our previous exact results for  the soft
$\beta$-functions and the AMSB solution to allow for the presence  of
gauge  singlet matter fields. In the usual formulation of the  NMSSM
(see for example \ref\kingwhite{S.~F.~King and P.~L.~White, \prd 52
(1995) 4183 }) it is easy to see that we would  have, in fact, $\sbar =
Z = 0$; for a non-zero $\cbar^i$  we obviously need a chiral superfield
which is a ``universal'' singlet (i.e.  invariant under both gauge and
global transformations). In the  standard gravity-mediated
supersymmetry-breaking scenario,  one may expect on rather general
grounds that $\cbar^i$ will  suffer gravity-induced  quadratic
divergences\ref\bapora{ J.~Bagger, E.~Poppitz and  L.~Randall, \npb 455
(1995) 59}  leading to contributions $\cbar^i \sim O(M_P M_{\rm
sparticle}^2)$,  and consequent destabilisation of the hierarchy. 
However there are frameworks where the gravitational tadpole  has a
magnitude that is phenomenologically acceptable  (or even desirable)
\ref\nillpol{
S.A.~Abel, \npb 480 (1996) 55\semi 
H.P.~Nilles and N.~Polonsky, \plb 412 (1997) 69\semi
C.~Panagiotakopoulos and A.~Pilaftsis, \prd 63 (2001) 055003\semi
A.~Dedes, C.~Hugonie, S.~Moretti and K.~Tamvakis, \prd 63 (2001) 055009
}.  We hope,
therefore, that our results  may prove of use in the  analysis of 
non-minimal versions of the MSSM. 

\bigskip\centerline{{\bf Acknowledgements}}\nobreak
 
This work was supported in part by the
Leverhulme Trust. RW was supported by a PPARC 
Research Studentship.

\listrefs
\bye